\documentclass[12pt]{article}
\usepackage{amssymb,amsmath,epsfig,multicol}
\usepackage{graphicx}

\def\today{\number\day\space\ifcase\month\or
January\or February\or March\or April\or May\or June\or July\or August\or
September\or October\or November\or December\fi \space\number\year}

\makeatletter

\def\@begintheorem#1#2{\trivlist \item[\hskip \labelsep{\sc #1\ #2.}]\sl}
\def\@opargbegintheorem#1#2#3{\trivlist
\item[\hskip \labelsep{\sc #1\ #2\ (#3).}]\sl}
\def\@endtheorem{\endtrivlist}

\def\@sect#1#2#3#4#5#6[#7]#8{\ifnum #2>\c@secnumdepth
\let\@svsec\@empty\else
\refstepcounter{#1}\edef\@svsec{\csname the#1\endcsname.\hskip 0.5em}\fi
\@tempskipa #5\relax
\ifdim \@tempskipa>\z@
\begingroup #6\relax
\@hangfrom{\hskip #3\relax\@svsec}{\interlinepenalty \@M #8\par} \endgroup
\csname #1mark\endcsname{#7}\addcontentsline {toc}{#1}{\ifnum
#2>\c@secnumdepth \else
\protect\numberline{\csname the#1\endcsname}\fi #7}\else
\def\@svsechd{#6\hskip #3\relax\@svsec #8.\csname #1mark\endcsname
{#7}\addcontentsline
{toc}{#1}{\ifnum #2>\c@secnumdepth \else \protect\numberline{\csname
the#1\endcsname}\fi #7}}\fi
\@xsect{#5}}

\newtheorem{theorem}{Theorem}[section]
\newtheorem{proposition}{Proposition}[section]

\newtheorem{definition}{Definition}[section]

\newenvironment{proof}{\removelastskip\vskip12pt plus 1pt \noindent \it
Proof.\quad\rm}{\hspace*{\fill}$\blacksquare$}
\newcommand{\be}{\begin{equation}}
\newcommand{\bT}{\begin{array}}
\newcommand{\eT}{\end{array}}
\newcommand{\de}{\end{equation}}
\newcommand{\ee}{\end{equation}}

\def\text#1{{\quad \hbox{#1} \quad}}

\begin{document}
\title{\bf Multiple solutions of steady MHD flow of dilatant fluids}
\author{
 {\textsc{Zakia Hammouch}}\footnote {Email adress: zakia.hammouch@u-picardie.fr} \\
{\footnotesize \it LAMFA, CNRS UMR 6140, Universit\'e de Picardie
Jules Verne
}\\
{\footnotesize \it Facult\'e de Math\'ematiques et d'Informatique,
33 rue Saint-Leu 80039 Amiens, France}
\date{}}
\maketitle
\noindent{\bf Abstract.\\}
In this paper we consider the problem of a steady MHD flow of a non-Newtonian power-law  and electrically conducting fluid in presence of an applied magnetic field. The boundary layer equations are solved in similarity form via the Lyapunov energy method, we show that this problem has an infinite number of positive global solutions. \\

\section{Introduction}
\setcounter{equation}{0}
\setcounter{theorem}{0}
\setcounter{lemma}{0}
\setcounter{remark}{0}
\setcounter{corollary}{0}
The study of non-Newtonian fluid flows has considerable interests, this is primarily because of the numerous applications in several engineering fields. Such processes are wire drawing, glass fiber and paper production, crystal growing, drawing of plastic sheets etc. For more details about the behavior in both steady and unsteady flow situations, together with mathematical models, we refer the reader to the books \cite{ast} by Astarita and Marucci, \cite{boh} by Bohme and the references therein. One particular non-Newtonian model which has been widely studied is the Ostwald-de Wael power-law model  \cite{acr}\cite{ece}, which  relies  the shear stress  to the strain rate $u_y $ by the expression
\begin{equation}\label{1}
 \tau_{xy} = k |u_{y}|^{n-1}u_{y},
\end{equation}
where $ k $ is a positive constant, and $ n  > 0 $ is called the power-law index .  The case $ n < 1 $ is referred to pseudo-plastic  or shear-thinning fluid, the case $ n > 1 $ is known as dilatant or shear-thickening fluid. The Newtonian fluid is a special case where the power-law index $ n $ is equal to one. In the present work we shall restrict our study to the case $n>1$.\\
The magnetohydrodynamics (MHD) flow problems find also applications in a large variety of physical, geophysical and 
industrial fields \cite{pav}. It is also interesting to study the flow of non-Newtonian fluids with externally imposed magnetic fields. To the author knowledge MHD flow of non-Newtonian fluids was first studied by Sarpkaya \cite{sar}. In \cite{sap} Sapunkov derived the equations describing the similarity solutions for the non-Newtonian flow when the external applied magnetic field varies as $x^{\frac{m-1}{2}}$, in presence of a pressure gradient, he used the method of series expansion. Later, Djuvic \cite{dju} employed a Crocco's variables to study the unsteady flow with exponentially  external velocity (in time). Recently, Liao \cite{lia} introduced a powerful technique (homotopy analysis) to give analytic solutions of MHD viscous flows of non-Newtonian fluids over a stretching sheet. \\
In this paper, we reconsider the steady two-dimensional laminar flow of an
   incompressible viscous electrically conducting dilatant fluid over
   a stretching flat plate with a power-law velocity distribution
   in the presence of a perpendicular magnetic field. Our interest in this work has been motivated by    the work of Chiam \cite{chiam}, who have considered the flow over an impermeable
    flat plate, for which similarity solutions were found via the Crocco transformation.
\section{Derivation of the model}
\setcounter{equation}{0}
\setcounter{theorem}{0}
\setcounter{lemma}{0}
\setcounter{remark}{0}
\setcounter{corollary}{0}
Consider a steady two-dimensional laminar flow of an incompressible
dilatant and electrically conducting fluid of density $\rho$, past a semi-infinite flat plate. Let $(x,y)$ be the Cartesian coordinates of any point in the flow domain, where $x-$axis is along the plate and $y-$axis is normal to it. Assume that a magnetic field $H(x)$, is applied normally to the plate.\\ The continuity and momentum equations can be simplified, within the boundary-layer approximation, into the following equations (see\cite{sap}\cite{chiam})
\begin{equation}\label{2}
u_{x} + v_{y} =0,
\end{equation}
\begin{equation}\label{3}
 uu_{x} + v u_{y}=\nu(|u_{y}|^{n-1}u_{y})_y+u_{e}{{u_{e}}_{x}}+\frac{\sigma \mu^{2} H^{2}}{\rho}(u_{e}-u).
\end{equation}
 Accompanied by the boundary conditions
\begin{equation}\label{4}
u(x,0) = U_{w}(x), \quad v(x,0) = V_{w}(x)\quad\mbox{and}\quad u(x,y) \to u_{e}(x)\quad
\mbox{ as } y \to \infty.
\end{equation}
\noindent Where the functions $ u $ and $ v $ are the velocity
components in the $ x $ and $ y $ directions respectively, $u_e(x)=U_{\infty}x^{m}$ is the free-stream velocity. The parameters $\nu,
n,\mu, \sigma$ and $H$ are the kinematic viscosity, the flow behavior
index, the magnetic permeability, the electrical conductivity of the
fluid, and  the magnetic field intensity respectively. The functions $ U_w(x) =
u_{w}x^m (u_{w}>0)$
 and  $V_{w}(x)=v_{w}x^{\frac{m(2n-1)-n}{n+1}}$ are the stretching
and the suction/injection velocities respectively.\\
In term of the stream-function ($\psi$ which satisfied \, $u(x,y)={\psi_y}(x,y)$ and \, $v(x,y)=-\psi_{x}(x,y)$), equations (\ref{2}),(\ref{3}) can be reduced to the single equation
\begin{equation}\label{5}
\psi_{y}\psi_{xy}-
\psi_{x}\psi_{yy} = \nu(|\psi_{yy}|^{n-1}\psi_{yy})_{y} + u_{e}{u_{e}}_{x} +
\frac{\sigma \mu^{2} H^{2}}{\rho}(u_{e}-\psi_{y}),
\end{equation}
subject to
\begin{equation}\label{5bis}
\psi_{y}(x,0)=u_w x^{m},\,\,\psi_{x}(x,0)=-v_{w}x^{\frac{m(2n-1)-n}{n+1}} \quad\mbox{and}\quad \psi_{y}(x,y)=U_\infty x^{m}\quad\mbox{as} \quad y \to \infty.
\end{equation}
 According to Sapunkov \cite{sap}, similarity solutions for problem (\ref{5}),(\ref{5bis}) exist only if the magnetic field has the following form\quad $H(x) \sim x^{\frac{m-1}{2}}$. \\
To look for similarity solutions we define the following
\begin{equation}\label{6}
  \eta := Ayx^{-a}\quad\mbox{and}\quad  \psi(x,y) := Bx^{b}f(\eta),
\end{equation}
\noindent where  $f$ is the transformed dimensionless stream
function and $\eta$ is the similarity variable.
 Thanks to (\ref{6}), the function $f$ satisfies the new boundary value problem
\begin{equation}\label{7}
\left\{\begin{array}{lll}
    (|f''|^{n-1}f'')'+a ff''+m(1-{f'}^{2})+M(1-f')= 0,\\
    \\
    f(0)=\alpha, \qquad f'(0)=\delta, \qquad   f'(\infty)=1,
    \end{array}\right.\end{equation}
if and only if 
 $$\displaystyle a = \frac{1+m(2n-1)}{n+1},\qquad b= \frac{1+m(n-2)}{n+1},\qquad a-b=m,$$
 and the parameters $A$  and $B$ satisfy
 $$AB= u_{\infty}\quad\mbox{and}\quad \nu B^{n-2}A^{2(n-1)}=1. $$
Where the primes denote differentiation with respect to $\eta$, the function $f'(\eta)$ denotes the normalized velocity and the parameters
$$\displaystyle M=\frac{\sigma \mu^{2} H_{0}^{2}(n+1)}{u_{\infty}\rho},\quad \displaystyle \alpha=-\frac{(n+1)v_{w}}{(m+1)(\nu u_{\infty}^{2n-1})^{\frac{1}{n+1}}}\quad \mbox{and}\quad \displaystyle \delta=\frac{u_w}{u_{\infty}},$$
are  respectively: The Hartmann number, the suction/injection and the stretching parameters.\\ Such problems have been investigated by several authors for example, Anderson et al. \cite{and}, Zhang et al. \cite{zh} and Kumari and Nath \cite{kum}. \\In the same context, Chiam \cite{chiam} studied  Problem (\ref{2})-(\ref{4}). To look for similarity solutions, he solved the following boundary value problem
  \begin{equation}\label{chiam}
\left\{\begin{array}{lll}
    n|f''|^{n-1}f'''+ ff''+\beta(1-{f'}^{2})+M(1-f')= 0,\\
    \\
    f(0)=0, \quad f'(0)=0, \quad  f'(\infty)=0.

\end{array}\right.
\end{equation}
Where $\beta=\frac{m(n+1)}{(2n-1)m+1}$. We aim here to stress that for $n\neq1$, equation (\ref{7})$_1$ can be degenerate at some point $\eta_{s}$ for which $f''(\eta_{s})=0$ (for more details see \cite{phd}) and then any solution of (\ref{7}) is not necessarily of $C^{3}(0,\infty)$. Hence equations (\ref{7})$_1$ and (\ref{chiam})$_1$ are not equivalent.\\
Let us notice that for the Newtonian case $(n=1)$, problem (\ref{7}) reduces to the Falkner-Skan flow in Magnetohydrodynamics, which has been studied by Hildyard \cite{hil}, Aly et al. \cite{aly1} and Hoernel \cite{david}.
The case $m=M=0$ leads to the generalized Blasius problem (see \cite{nach}).
While the case  $m=-M$, by a suitable scaling,  is referred to the mixed convection of a non-Newtonian fluid in a porous medium (see for example \cite{mixed1}). We note also that in absence of the magnetic field, problem (\ref{7}) is simplified to the Falkner-Skan flow for non-Newtonian fluids. A complete study on this subject is given in \cite{denier} by Denier and Dabrowski.\\
Very recently, Aly et al. \cite{aly1} reported a theoretical and numerical investigations
on the existence of solutions to problem (\ref{7}) for Newtonian fluids $(n=1)$, say
\begin{equation}\label{aly1}
\left\{\begin{array}{lll}
    f'''+\frac{m+1}{2} ff''+m(1-{f'}^{2})+M(1-f') = 0,\\
    \\
    f(0)=\alpha \geq 0, \quad f'(0)=\delta, \quad  f''(0)=\gamma.

\end{array}\right.
\end{equation}
They showed that problem (\ref{aly1}) has multiple solutions for any $\delta\in(0,\Gamma)$ and $\gamma\in\mathbb{R}$ satisfying
\begin{equation}\label{aly2}
\gamma^{2}\leq\frac{2m}{3}\delta^{3}+M\delta^{2}-2(M+m)\delta, 
\end{equation}
where  $\displaystyle \Gamma=-\frac{3M}{4m}\left[1+\sqrt{1+\frac{16m}{3M^{2}}(m+M)}\right]>1$. In the present work, we aim to extend their results to the non-Newtonian dilatant fluids $(n>1)$, by using a condition on $\gamma$ which is different from (\ref{aly2}) and without any restriction on the parameter $\delta$. 

 \section{Non-uniqueness of solutions}
\setcounter{equation}{0}
\setcounter{theorem}{0}
\setcounter{lemma}{0}
\setcounter{remark}{0}
\setcounter{corollary}{0}
Guided by the analysis of \cite{bzaa},\cite{phd} and \cite{zamp}, we aim to prove the existence of solutions to problem (\ref{7}), for related values of the parameters $m, M, n, \alpha, \delta$ and $\gamma$. This result will be established by mean of the so-called shooting method,  the  boundary value problem (\ref{7}) is then converted into the  following initial value problem 
\begin{equation}\label{8}
\left\{\begin{array}{lll}
    (|f''|^{n-1}f'')'+a ff''+m(1-{f'}^{2})+M(1-f') = 0,\\
    \\
    f(0)=\alpha, \quad f'(0)=\delta, \quad  f''(0)=\gamma.

\end{array}\right.
\end{equation}
Where the real number $\gamma$ is the shooting parameter.\\
The initial value problem (\ref{8}) can be transformed into the equivalent first order ordinary differential system
\begin{equation}\label{9}
\left\{\begin{array}{lll}
f'=g,\\
\\
g'=|h|^{\frac{1-n}{n}}h,\\
\\
h'=-a f |h|-m(1-g^{2})-M(1-g),\\
\end{array}\right.
\end{equation}
with the conditions
\begin{equation}\label{10}
 f(0)=\alpha, \quad g(0)=\delta, \quad  h(0)=|\gamma|^{n-1}\gamma.
\end{equation}
By the classical theory of ordinary differential equations, problem (\ref{9}),(\ref{10}) has a unique local (maximal) solution for every $\gamma\neq0$. Let $f_{\gamma}$ denotes this solution and $(0,\eta_{\gamma}),$ $\eta_{\gamma}\leq\infty,$ denotes its maximal interval of existence. The main task now is to show how existence of solutions depends on $\gamma$.\\ The local solution $f_\gamma$ satisfies the following  
 \begin{equation}\label{11}
|{f''}_{\gamma}|^{n-1}{f''_{\gamma}}+a {f'}_{\gamma}f_{\gamma}
-M(f_{\gamma}+\alpha)=|\gamma|^{n-1}\gamma+a \alpha \delta -(M+m)\eta +
(a +m){\int_{0}}^{\eta}{f'_{\gamma}}^{2}(\tau)d\tau.
\end{equation}
Equation (\ref{11}) will be used for proving the main results.
 \begin{definition}
 A function $f_{\gamma}$ is said to be a solution to (\ref{8}) if $f\in C^{2}(0,\infty)$, $|f_{\gamma}''|^{n-1}f_{\gamma}'' \in C^{1}(0,\infty)$ and satisfies
 $$\lim_{\eta\rightarrow \infty} f'_{\gamma}(\eta)=1 \,\,\, \mbox{(i)}\,\,\,\,\mbox{and}\,\,\, \lim_{\eta\rightarrow \infty}\,\,\,
{f''_\gamma}(\eta)=0\,\,\mbox{(ii)}$$
 \end{definition}
\subsection{Suction/Injection flows ($\alpha \in \mathbb{R}$)}
\begin{theorem}\label{th1}
Assume $\alpha\in \mathbb{R}$, $\delta>0$, $M>0$ , $n > 1 $ \,and\, $-\frac{1}{3n}<m<-M$. For any $\gamma$ satisfying $$|\gamma|^{n-1}\gamma >-a \alpha \delta \,\, (\star),$$
 problem (\ref{8}) admits a global unbounded solution.
\end{theorem}

\begin{proof}{\rm
From a physical point of view, it is more convenient to prove the result for  the cases $\alpha\geq 0$ (suction) $\alpha < 0$ (injection) separately.\\
We have to show that $f_{\gamma}$ is a
positive monotonic increasing function on $(0,\eta_{\gamma})$,
globally defined and going to infinity with $\eta$.
For this sake we define the Lyapunov Energy function by
\begin{equation}\label{12}
V(\eta)=\frac{1}{n+1}|f''|^{n+1}-\frac{m}{3}f'^{3}-\frac{M}{2}f'^{2}+(M+m)f'.
\end{equation}
Which satisfies  $$V'(\eta)=-a ff''^{2}.$$ Then $V$ is monotonic decreasing on $(0,\eta_\gamma)$. On the other hand, from equation (\ref{11}) and condition $(\star)$ we see that ${f_{\gamma}}'$ and $f_{\gamma}$ are positive on $(0,\eta_\gamma)$ as long as $f_{\gamma}$ exists. Using the Lyapunov function $V$ we see that ${f_{\gamma}}'$ and ${f_{\gamma}}''$ are bounded, since $V$  is bounded from below by $\frac{3M+4m}{6}$. If $f_\gamma$ were also bounded, say ${f \rightarrow}_{\eta\infty} L$ with $L \in (0,\infty)$ (since $f_\gamma$ is positive). Then $ {f_\gamma}'(\eta) \rightarrow_{\eta\infty} 0$ which implies that $ f_\gamma''(\eta_{k})\rightarrow_{k\infty} 0$, where $(\eta_{k})_{k\geq0}$ is a sequence tending to infinity with $k$. Using again (\ref{11}) to deduce
$$|{f}_{\gamma}''(\eta_k)|^{n-1}f_{\gamma}''(\eta_k)+ a {f}_{\gamma}'(\eta_k)f_{\gamma}(\eta_k)
 = - M(f_{\gamma}(\eta_k)+\alpha)+|\gamma|^{n-1}\gamma+a \alpha \delta -$$
$$(M+m)\eta_k +
(a +m){\int_{0}}^{\eta_k}{f'_{\gamma}}^{2}(\tau)d\tau.$$
 Letting $k \rightarrow \infty$, the right hand side goes to zero while the left hand side goes to minus infinity, which is impossible. then $f_{\gamma}$ is a global unbounded solution to (\ref{8}).\\  
From the above $f_\gamma'$ and $f_\gamma''$ are bounded and $f_{\gamma}'$ is monotonic increasing on \,$(\eta_1,\infty)$,\, for \,$\eta_1$\ large enough. Then there exists $l>0$
such that $\lim_{\eta\rightarrow \infty} f_{\gamma}'(\eta)=l,$ and there exits a sequence $(\zeta_k)_k$, tending to infinity with $k$ such that $\lim_{k\rightarrow \infty}f_{\gamma}''(\zeta_k)=0$. Making recourse to the Lyapunov function $V$ we get 
$\lim_{\eta\rightarrow \infty}
f_{\gamma}''(\zeta_k)=0$.\\ Assume now that $f_\gamma''$ is not monotonic on any interval $[\eta_2,\infty)$. Then, there exists a sequence $(\tau_k)_k$ going to infinity with $k$ such that:\\
$\bullet$\, $ (|f_\gamma''|^{n-1}f_\gamma'')'(\tau_{k})=0,$\\
$\bullet$\, $|f_\gamma''|^{n-1}f_\gamma''(\tau_{2k}) \quad \mbox{is a local minimum},$\\
$\bullet$\, $ |f_\gamma''|^{n-1}f_\gamma''(\tau_{2k+1}) \quad \mbox{is a local maximum}.$\\
From (\ref{8})$_1$, we have
 $$f_{\gamma }''(\tau_{k})= -\frac{m(1-f_{\gamma}'^{2}(\tau_{k}))+M(1-f_{\gamma}'(\tau_k))}{a f_{\gamma}(\tau_{k})}.$$
Since $f_\gamma' $ is bounded and $f_\gamma $ goes to infinity with $\eta_k$, we get easily from the above that $f_{\gamma}''$ goes to zero with $\eta$.\\
 Now we show that $f_{\gamma}$ satisfies (i). Recall that $f_{\gamma}'$ is a positive bounded function then $f_{\gamma}'\rightarrow l$ with  $l\in (\alpha ,\infty)$. At infinity we have $f_\gamma  \sim \eta l$  and from identity (\ref{11}) we get
$$|f_\gamma''|^{n-1}f_{\gamma}'' \sim \eta[ml^{2}+Ml-(M+m)]+ o(1)$$ as $\eta$ approaches infinity, 
this leaves only the possibility that $l$ is either $1$ or $-\frac{M}{m}-1$, thanks to the positivity of $f_\gamma'$ we deduce that $l=1$. \\
To finish we show the result for $\alpha <0$. In such case, the function $f_\gamma$ is negative on a small neighborhood of zero. According to (\ref{11}) $f_\gamma$ cannot have a local maximum, then two possibilities arise:\\
$\bullet$\, $\mbox{Either}\qquad f_{\gamma} < 0 \,\, \forall \, \eta\in(0,\eta_\gamma)$\\
$\bullet$\, $\mbox{Or}\,\quad\qquad \exists \, \eta_{\star } \quad \mbox{such that}\quad f_{\gamma} < 0 \quad\mbox{on} \quad (0,\eta_\star), \quad f_{\gamma}(\eta_\star )=0\qquad\mbox{and}\quad f_{\gamma} > 0\quad \forall  \eta > \eta_\star.$\\
Assume that the first assertion holds, then $\alpha > f_{\gamma}(\infty) \leq 0$ and $f_{\gamma}'(\infty)=0$, $f'$ being positive we use again (\ref{11}) to get that $f_{\gamma}''$ is positive. A contradiction. Then, $f_\gamma$ has exactly one zero $\eta_\star $. We define the shifted function $h$ by :
$$\eta \longmapsto h(\eta)= f_{\gamma}(\eta+\eta_{\star }),$$
which satisfies $h(0)= 0$, $h'(0)= \delta$ and $h''(0)> 0$, and we use the above  analysis to conclude that $h$ is an unbounded global solution to (\ref{8}).}
\end{proof}
\subsection{Reversed flows $(\delta <0)$}
Now we pay attention to the case of reversed flows $(\delta<0).$ First, we show that the shooting parameter has to be positive. 
\begin{proposition}
Let $f_{\gamma}$ be a solution to (\ref{8}) with $m\in(-\frac{1}{3n},-M),$ $\alpha < 0$, $\delta < 0$ and  $\gamma \leq 0$, then the  condition (i) is failed.
\end{proposition}
\begin{proof}{\rm
Let $\delta < 0$, if $\gamma \leq 0$ then ${f_\gamma}''$ is negative on some $(0,\eta_{0})$, for $\eta_0$ small, and equation (\ref{8}) can be written as
$$ ({f_\gamma}''e^{F})' = - \frac{e^{F}}{n}|f_{\gamma}''|^{1-n} \left[ m(1-f_{\gamma}'^{2})+M (1-f_{\gamma}')\right] ,$$
where $\displaystyle F(\eta) = \frac{a}{n} {\int_0}^{\eta}f_{\gamma}|{f_{\gamma }}''|^{1-n}d\tau$. From this we see that $\eta \longmapsto f'' e^{F}$ decreases and then $f_{\gamma}''(\eta) \leq 0$ for all $\eta \in (0,\eta_{\gamma})$. It follows that
$f_{\gamma}'$ is decreasing on $(0,\eta_\gamma)$ and then the condition (i) could not be satisfied.}
\end{proof} 
\begin{theorem}\label{th2} Let $ \delta < 0, \alpha > 0 $ and $m \in (-\frac{1}{3n},-M).$ For any $
\gamma > 0 $ satisfying $$\alpha \gamma^{n} - \frac{1}{2}\delta^2\gamma^{n-1} + a \alpha^2\delta-\frac{M}{2}\alpha^{2} >0 \,\,(\star\star),$$ problem (\ref{8}) has a global unbounded solution.
\end{theorem}
\begin{proof}{\rm
Let $f_{\gamma}$ be the local solution of (\ref{8}), define the auxiliary function
\begin{equation}\label{13}
G(\eta) = f_\gamma  {f_\gamma}'' |f_{\gamma }''|^{n-1}-\frac{1}{2}f_{\gamma }'^{2}|f_{\gamma}''|^{n-1}+a f_{\gamma}^{2}f_{\gamma}'-\frac{M}{2}f_{\gamma}^{2},
\end{equation}
which satisfies
\begin{equation}\label{14}
G'(\eta) = -(m+M)f_\gamma + \left[2a+m+\frac{(n-1)a}{2n}\right] f_{\gamma }'^{2}f_{\gamma}+\frac{n-1}{2n}f_{\gamma}'^{2}{f_{\gamma}''}^{-1}\left[ m(1-{f_{\gamma}'}^{2})+M(1-f_{\gamma}')\right] ,
\end{equation}
and $G(0)> 0$.\\ Since $f_{\gamma}' < 0$, the function $f_\gamma$ is negative on a small neighborhood of zero. Assume that there exists $\eta_{1}\in (0,\infty)$ such that $$f_{\gamma}(\eta)> 0,\quad f_{\gamma}'< 0\qquad \forall \eta\in[0,\eta_1)\quad \mbox{and} \quad f_{\gamma}(\eta_{1})=0.$$
Hence $G$ is a monotonic nondecreasing function on $(0,\eta_1)$ and then $G(\eta_1)\leq 0$. Then $G(\eta)\leq 0$ for all $\eta\in(0,\eta_1)$, in particular $G(0)\leq 0$, which is a contradiction with $(\star\star)$. Therefore we have :\\
$\bullet$ $\mbox{Either}\quad f_{\gamma}>0\quad\mbox{and}\quad f_{\gamma}'\leq 0\quad{\forall}
\eta\geq 0$\\
$\bullet$ $\mbox{Or}\quad \exists \eta_{2}>0\, : \quad f_{\gamma}>0, \quad f_{\gamma}'<0 \quad \forall \eta \in (0,\eta_2),\quad f_{\gamma}'(\eta_2)=0\quad \mbox{and}\quad f_{\gamma}(\eta_2) \quad \mbox{is a local maximum}.$\\
Assume that the first assertion holds, then $f_{\gamma}$ has a finit limit at infinity, say $L\in (0,\infty)$ and there exists a sequence $(\chi_{k})_k\geq0$ tending to infinity with $k$ such that $f_{\gamma}'(\chi_{k})$ goes to zero at infinity. If $f_{\gamma}'$ is monotonic (resp. non-monotonic on any interval $(\eta,\infty)$) we get $f_{\gamma}'$ goes to zero at infinity and then  $f_{\gamma}''(\delta_{k})$ goes to zero
 at infinity for a sequence $(\delta_k)_{k\geq0}$ going to infinity with $k$ (resp. $f_{\gamma}''(\delta_k)=0$ and $f_{\gamma}'(\delta_{k})$ goes to zero at infinity). Because $G(0)< G(\delta_{k})$, we obtain a contradiction by taking the limit as $k$ goes to infinity. \\Now, we claim that the function $f_{\gamma}$ cannot have a local maximum after $\eta_{2}$. Actually, assume  there exists $\eta_{3}>\eta_{2}$ such that $f_{\gamma}(\eta_{3})$ is a local maximum. At this point the function $G$ takes a negative value and satisfies $ 
G(\eta_{3}) \geq G(0)$ a contradiction. Since $f_{\gamma}$ is monotonic increasing after $\eta_{2}$ we deduce as the above that is a global solution.\\ Next, we argue as in the proof of Theorem. \ref{th1} to show that $f_{\gamma}$ is unbounded at infinity and satisfies (i) and (ii).}
\end{proof}  
\subsection{Flow with large initial velocity ($\delta \gg 1$)}
In this subsection, we construct asymptotic solutions to problem (\ref{8}) when the real $\delta$ is very large. Adopting the method used in \cite{aly2} by Aly et al., we assume that such solutions can be written under the following form
$$ f(\eta)=\eta+\xi^{r}g(t),\quad \mbox{where} \quad t=\xi^{s}\eta, \quad \xi=\delta-1 \quad \mbox{and} \quad r,s \in \mathbb{R}.$$
Then problem (\ref{8}) reads
\begin{equation}\label{15}
\left\{\begin{array}{lll}
    \xi^{(r+2s)(n-2)}(|g''|^{n-1}g'')'+a  \eta   \xi^{-r}g''+a g g''-(2m+M)\xi^{-(r+s)}-m {g'}^{2} = 0,\\
    \\
    g(0)=\alpha \xi^{-r}, \quad g'(0)=\xi^{1-(r+s)}, \quad  g'(\infty)=0.
\end{array}\right.
\end{equation}
Setting $r=\frac{2n-1}{n+1}$ and $s=\frac{2-n}{n+1}$, ensures that the highest derivative remains present in the resulting problem.\\
As $\xi$ goes to infinity, we deduce
\begin{equation}\label{16}
\left\{\begin{array}{lll}
    (|g''|^{n-1}g'')'+a gg''+m{g'}^{2}= 0,\\
    \\
    g(0)=0, \quad g'(0)=1, \quad  g'(\infty)=0.

\end{array}\right.
\end{equation}
Problem (\ref{16}) describes the steady free convection flow of a non-Newtonian power-law fluid over a stretching flat plate embedded in a porous medium. In \cite{phd}, it was shown that for $m\in(-\frac{1}{3n},0)$ any local solution $g$, whith positive values of $\tau$ $(\tau=g''(0))$, is global and satisfies the following asymptotic behaviour
$$g(t) \sim t^{\frac{1+m(2n-1)}{1+m(n-2)}},\qquad \mbox{as} \quad  t\to \infty.$$
Consequently, a solution $f$ for  positive $\gamma$ and  large $\delta $ (if it exists), may have the following large $\eta$-behaviour
$$f(\eta) \sim \eta\left[1+(\delta-1)^{\frac{1}{1+m(n-2)}}\eta^{\frac{m(n-1)+2}{1+m(n-2)}}\right] .$$
\section{Concluding remarks}
 Based on the similarity transformation approach, the boundary layer equations for flows of purely viscous non-Newtonian dilatant and electrically conducting fluids are investigated. Using a shooting argument, it is shown that the relevant problem admits an infinite number of solutions (\cite{magyari}\cite{liao}\cite{riley} and \cite{rob}), this is due to the arbitrariness of the shooting parameter $\gamma$. 
 From a physical point of view, we underline  that $\gamma=f''(0)$  originates from the local skin friction coefficient $C_{f_{x}},$  and the local Reynolds number $\displaystyle Re_x=\frac{({u_{w}x^{m}})^{2-n}x^{n}}{\nu k}$  via the the formula    
$$\displaystyle C_{f_{x}} {Re_x}^{\frac{1}{n+1}}= 2{\left(\frac{a}{n}\right)}^{\frac{1}{n+1}}|\gamma|^{n-1}\gamma.$$
In conclusion, we may expect that the solutions determined above are physically acceptable. However, only experiments are able to prove their physical existence.

\section*{Acknowledgements}
I would like to thank the anonymous referee for his constructive suggestions, which have improved the earlier version of this work.

\end{document}